\documentclass[12pt,preprint]{aastex}

\def\etal{{\it et al.\ }}

\def\cal#1{{\cal #1}}

\def\m@th{\mathsurround=0pt}
\def\n@space{\nulldelimiterspace=0pt \m@th}
\def\biggg#1{{\mbox{$\left#1\vbox to 20.5pt{}\right.\n@space$}}}

\def\beginenum{\begin{enumerate}}
\def\endenum{\end{enumerate}}
\def\bitem{\begin{itemize}}
\def\eitem{\end{itemize}}
\def\bray{\begin{array}}
\def\eray{\end{array}}
\def\begindoc{\begin{document}}
\def\enddoc{\end{document}}
\def\bc{\begin{center}}
\def\ec{\end{center}}

\begin{document}
\title{Comments on "The Coronal Heating Paradox" \\
by M.J. Aschwanden, A. Winebarger, D. Tsiklauri and H. Peter
[2007, \apj, 659, 1673]}
\author{Swadesh M. Mahajan\altaffilmark{1}}
\affil{Institute for Fusion Studies, The University of Texas at
Austin, Austin, Texas 78712}
\author{and \\
\vspace{0.3cm}
Nana L. Shatashvili\altaffilmark{2}} \affil{Faculty
of Exact and Natural Sciences, I. Javakhishvili Tbilisi State
University, Tbilisi 0128, Georgia\\
Andronikashvili Institute of Physics, Tbilisi 0177, Georgia}
\altaffiltext{1}{\small Electronic mail: \
mahajan@mail.utexas.edu}
\altaffiltext{2}{\small Electronic mail:
\ shatash@ictp.it \hskip 0.3cm shatashvili@yahoo.com}

\clearpage

\begin{abstract}
We point out the priority of our paper \cite{MMNS2} over
\cite{AWTP} in introducing the term "Formation and primary heating
of the solar corona" working out explicit models (theory as well
as simulation) for coronal structure formation and heating. On
analyzing the \cite{AWTP} scenario of coronal heating process
(shifted to the chromospheric heating) we stress, that for
efficient loop formation, the primary upflows of plasma in
chromosphere/transition region should be relatively cold and fast
(as opposed to hot). It is during trapping and accumulation in
closed field structures, that the flows thermalize (due to the
dissipation of the short scale flow energy) leading to a bright
and hot coronal structure. The formation and primary heating of a
closed coronal structure (loop at the end) are simultaneous and a
process like the "filling of the {\it empty} coronal loop by hot
upflows" is purely speculative and totally unlikely.
\end{abstract}


\section*{}
\clearpage

\bigskip

In a recently published paper,  "The Coronal Heating Paradox" by
M.J. Aschwanden, A. Winebarger, D. Tsiklauri and H. Peter [AWTP
(2007)], it was pointed out:

A1) that observations show no evidence for local heating in the
solar corona, but rather for heating below the corona in the
transition region (TR) and upper chromosphere, with subsequent
evaporation as known in flares,

A2) that  the phrase "coronal heating problem" is therefore a
paradoxical misnomer for what should rather be addressed as the
"chromospheric heating problem" and "coronal loop filling
process".

A3) that the hot temperature of the solar corona is generated by a
primary heating process located in the solar transition region or
upper chromosphere.

Before we present  a critical analysis of the "proper" use of
observational constraints that they invoke to arrive at their
conclusions, we would like to draw the authors' as well as the
community's attention to our paper "Formation and Primary Heating
of the Solar Corona - Theory and Simulation", Mahajan {\it et
al.}, 2001, Phys. Plasmas, {\bf 8}, 1340 (MMNS) published six
years ago. A short summary of the substantive aspects of MMNS
follows:

M1) MMNS first "invokes the term "primary heating of the solar
corona". In fact, it appears in the very title of the paper.

M2) discusses in great detail possible scenarios of closed coronal
structure formation and heating.

M3) MMNS is based on an integrated Magneto--Fluid model that
accords full treatment to the Velocity fields associated with the
directed plasma motion; model was developed to investigate the
dynamics of coronal structures formation.

M4) One of the principal objects of MMNS was  to suggest and
investigate the notion that the interaction of the fluid and the
magnetic aspects of plasma may be a crucial element in creating so
much diversity in the solar atmosphere.

M5) It was shown that the structures which comprise the solar
corona can be created by plasma flows observed near the Sun's
surface --- the primary heating of these structures is caused by
the viscous dissipation of the flow kinetic energy. The structure
formation and primary heating are simultaneous -- when the coronal
loop appears it is already hot.

M6) Some detail: we proposed that the high speed ($\leq
300$\,km/sec) streaming of plasma up through the height of the so
called base of the corona (observed by TRACE, and to some extent
by earlier space observations of the Sun), is the primary source
of hot material that makes up the corona. The primary plasma flow
becomes heated to $10^6$\,K or more as it is slowed down (up at
the coronal base height) by passage through a shock front.
$300$\,km/s  is adequate to be thermalized to  $(1-4)\cdot
10^6$\,K  in a shock transition. We also elaborated the model and
suggested that there are at least two major stages in the "heating
of the corona" (more accurately of a coronal structure): 1. A fast
primary heating period (as observations show and simulations
demonstrate) simultaneous with the creation of the hot coronal
structure, and 2. An auxiliary or a supporting period needed to
make up the losses and sustain the hot structure for a longer
time.

M7) More detail: In the MMNS model, the first stage plasma
"up-flows" (primary flows) are cold when entering the closed
magnetic field regions but they provide the needed energy and
material for the formation of the coronal structure. The second
stage could be fuelled by a variety of mechanisms like
magneto--fluid coupling \cite{MMNS1,MMNS2}, wave-dissipation, and
reconnections. The TRACE observations, suggesting $\leq
300$\,km/sec upward flows of plasma were the inspiration for the
basic MMNS idea [see observational data in
\cite{MMNS2,osym1,osym2,mnsy,RD,msms} and references therein]. We
stress that the exact value of the flow--velocity is not crucial;
to explain the formation and primary heating era of the life of
the hot closed coronal structure, the flow, however, must be
supersonic for the chromospheric temperatures at the relevant
heights. Very hot flows will not be, in general, supersonic and
the shock formation could not dissipate their kinetic energy to
heat. In this case we believe that the magneto--fluid mechanism
may do the trick \cite{MMNS2}. If the flows are subsonic and cold
then this very mechanism will be still operative but the end
product (coronal loop) will be cooler than 1MK; then the
additional heating mechanisms should be imposed to get the bright
loop. Essential is, that in all cases plasma is accumulated and
one gets {\it an overdense loop with specific temperature
different from that of primary flow}.

M8) For the MMNS paper and our later work in the field, we have
greatly  benefitted from various papers (including observational)
bearing the names from the AWTP list. We have learnt:

i) that the solar corona is a highly dynamic arena replete with
multiple--scale spatiotemporal structures \cite{A1,A2} indicating
a significant role for dynamical processes in coronal heating
\cite{klimchuk,warren}. Observations suggest that there are
strongly separated scales both in time and space in the solar
atmosphere [see e.g. \cite{KB,schrijver}, also \cite{klimchuk2}
and references therein]. Loops at different temperatures exist in
the same general region and may be co--located to within their
measured diameters. The large line shifts, or high velocities
($\pm 50 -– 100$\,km/s , or even $200 –- 300$\,km/s), are most
common at transition region temperatures $T \leq 5\cdot 10^5$\,K,
and seldom appear at $1$\,MK. Cool loops in active regions show
temporal variability. Characteristic times for the changes may
vary. A loop system may be quiescent for a long time with
individual loops living for several hours (recall our 2nd era of
quasi-equilibrium in the life of the closed coronal structure).
Quiescent periods may be followed by rapid activity (loops are
"turned on"/disappear in $\leq 10-40$\,min). Flows are found
within loops. The time variable emission over a full range in
temperatures in a volume filled with transient loops points to a
close connection between regions of various temperature
structures, at least in the range from $10^4$\,K to $1.5\cdot
10^6$\,K".

ii) from the MMNs perspective, a major new advance is the
discovery that strong flows are found everywhere --- in the
subcoronal (chromosphere/transition region) as well as in the
coronal regions (see e.g. [Kjeldseth-Moe \& Brekke 1998; Schrijver
\etal 1999; Winebarger \& DeLuca \& Golub 2001; Wilhelm 2001;
Aschwanden \etal 2001a; Aschwanden \etal 2001b; Seaton \etal 2001;
Winebarger \etal 2002] and references therein). Equally important
is the growing belief and realization that the plasma flows may
complement the abilities of the magnetic field in the creation of
the amazing richness observed in the coronal structures
\cite{MMNS1,MMNS2,channel1,channel2}.

iii) It stands to reason, then, that  one should investigate the
single hot closed coronal structure formation and primary heating
process rather than the heating of the entire corona (see the next
paragraph). Depending on the boundary and initial conditions taken
for chromosphere/TR, one will find  different dynamics of the
formation and different final parameters of given loop. We cite
here our conclusion from \cite{mnsy}, {\it The coronal heating
problem, ..., is shifted to the problem of the dynamic
energization of the chromosphere. ... the coronal heating problem
may only be solved by including processes (including the flow
dynamics) in the chromosphere and the transition region.} The
challenge, therefore, was to develop a semi--steady state theory
of flow generation in the chromosphere that we performed later
based on the suggested and explored magneto-fluid coupling
\cite{mnsy,RD,msms}. The correct theoretical model should explain
the global dynamics of given solar atmosphere region.

M9) Thus, at any quasi-equilibrium stage of the accelerating
plasma flow [the acceleration scenario could be one of many], the
nascent intermittent flows will blend and interact with
pre--existing closed field structures on varying scales. "New"
flows could be trapped by other structures with strong/weak
magnetic fields and participate in creating different dynamical
scenarios (when dissipation is present) leading to:

1) Heating of a new structure of the finely structured atmosphere
[see \cite{MMNS1,MMNS2}]. 2) Explosive events/prominences/CME
eruption [see \cite{osym1,osym2,mnsy}]. 3) Creation of a dynamic
escape channel (providing important clues toward the creation of
the solar wind [see \cite{channel1,channel2}]). 4) Instabilities,
and wave-generation could also be triggered.

In this context we can now stress that our conjecture is rather
general, and goes beyond just the heating issue (and even
formation issue) for a specific coronal structure:

M10) Formation and primary heating of coronal structures as well
as the more violent events (flares, erupting prominences and CMEs\
) are the expressions of different aspects of the same general
global dynamics that operates in a given coronal region.

M11) The plasma flows, the source of both the particles and energy
(part of which is converted to heat), interacting with the
magnetic field, become determinants of a wide variety of plasma
states comprising the observed coronal structures. The dissipation
of short--scale component of the velocity field may provide a
primary (during very formation) and a secondary (supporting)
heating for the coronal structure (closed, open).

M12) We repeat: the formation and heating are contemporaneous --
primary flows are trapped, accumulated and a part of their kinetic
energy dissipates during their trapping period. It is the initial
and boundary conditions that define the characteristics of a given
structure (coronal structure $T_c\gg T_{0flow}\geq 2eV$).

M13) When studying  the formation dynamics of a  closed coronal structure,
one has to distinguish between, and model 2 distinct eras  \cite{MMNS2}:

i) A hectic dynamic period when it acquires particles and energy
(accumulation $+$ primary heating). Naturally the description must
be time dependent.

ii) Quasistationary period when it "shines" as a bright, high
temperature object.

M14) In equilibrium each coronal structure has a nearly constant
$T$, but different structures have different characteristic $T$-s,
i.e., bright corona seen as a single entity will have considerable
$T$--variation.

M15) The term: "primary heating" can not be used separately from
the term "formation of the structure" when speaking of the heating
of fineley structured dynamical solar atmosphere. Also the correct
phrases  for chromopsheric heating  should read as: "primary
heating of the chromospheric part of solar atmosphere loop";
"creation of coronal base of the same loop" and "formation of
solar atmosphere loop".

M16) The heating of Coronal holes (dynamical rather than given
steady structures) is strictly linked to the bulk acceleration of
plasma there, as well as to the dynamical magnetic field openings,
and, thus, has to be treated differently. We would just stress
that heating mechanism that we suggest clearly explains the
difference in the final temperatures of ions and electrons in CH
-- the viscous heating of ions is much stronger than electrons.

\bigskip

We are somewhat puzzled and highly disappointed that the authors
of \cite{AWTP} have chosen to completely ignore our considerable
published work in this  general field; a field that we have
investigated over several years and in which we have even
introduced some of the nomenclature and vocabulary they use. We
have even communicated with at least two of the authors (sending
our papers etc.) and two of our papers were cited by one of them
earlier. This time, while using our terminology but for a
different scenario, they do not show adequate respect to our
rather encompassing and detailed investigations that produced,
inter alia, a scenario for the formation and heating of the
coronal structures.

We now go back and critique some of the arguments of AWTP (2007):

C1) They claim that "the observed overdensity of heated loops with
respect to background corona need to be explained by every
mechanism". And that "this fact can not be explained by local
heating mechanisms (like wave-heating or magnetic reconnections)".

Here  the authors believe that there can not exist cool solar
atmosphere loops that are denser than background corona, though
observations do show that such loops exist (see cited references
above). We have clearly shown (points M6, M7, M11-14) that
explanation of overdensity is rather straightforward, in fact,
automatic, when one invokes primary upflows but follows a scenario
different from AWTP (2007) . At the same time heating is neither
local nor external -- it is due to the proper understanding of the
magneto--fluid coupling when the primary flows (point M4) are
introduced.

C2) According to AWTP (2007) (section 3.2) "the electron density
in coronal loops can be only  significantly enhanced by upflows
of additional plasma from the loop footpoints, which requires a
secondary heating process in the upper chromosphere or lower
transition region".

We believe that here  the terms {\it primary heating} and {\it
secondary heating} are not being used carefully; both terms seem
to refer to the same process of chromospheric heating of
additional plasma (upflows).  Everywhere before and after this
statement, the authors use "primary heating" only for the
chromospheric heating of the flows that will later fill the
putative empty loops. One could imagine that they need to support
the hot chromospheric upflows heating process for later times (to
have continuous filling of loops) or for some other reason (e.g.
{\it chromospheric evaporation} -- see C6) below) . If so, then,
this supporting stage could be called the "secondary heating".
Unfortunately there is no indication given in the AWTP (2007)
about such supporting stage [we could only guess about such stage
when taking their statement presented in C6) -- see {\it excess
heating} term there]. On the contrary, their conclusion given at
the end of section 3.4 \ "this cycle of hot upflows and cool
downflows clearly points to the transition region as primary
heating source" explicitly says that they consider the
chromospheric heating to be primary.

C3) AWTP (2007) implies that the loop footpoints are above the
chromosphere/TR. Our calculations show  explicitly that  the
magneto--fluid interaction starts to operate immediately as the
flows enter the closed magnetic field region and interact with
them. The loop footpoints lie far below the coronal base (due to
the fact that the magnetic field emerges from the photosphere);
the coronal base (CB) is the place where the temperature
corresponding to the hot corona is reached; the  CB and the "loop
footpoint" are, generally, quite apart. The  loop is a composite
atmospheric structure; it has several different parts (see point
M15).

C4) AWTP (2007) states the following: "every primary heating
mechanism in the corona is not able to explain the observed
overdensity or emission measure excess in hot coronal loops,
unless chromospheric evaporation occurs".

Postulating  the {\it chromospheric evaporation} alone as
responsible for overdensity without the benefit of an implicit
theoretical model or a dynamical simulation is highly speculative
and lacks content. To support their argument ({\it for
chromospheric heating with subsequent coronal filling}) they give
examples from {\it observations of hot upflows from the
chromosphere}; this challenges the very idea of a closed coronal
structure -- this atmospheric structure, surely, has varying scale
footpoints much below the hot CB and it is unlikely that there
preexists some "empty" vessel like a "loop" which is waiting to be
filled with hot plasma upflows.

C5) How does this filling take place? The authors do not explain
this. They, of course, ignore the magneto--fluid coupling. We
showed earlier (see discussion above and MMNS) that the given
upflows are already thermalized (and hence slowed down) and
accumulated at the so called "coronal base" that is dynamically
created and not given initially (!!) -- when one observes the
hot/bright coronal structure the job by chromospheric cold and
relatively fast upflows is already done (reminding the reader that
cooler overdense loops have the right to exist as well).

C6) The essence of the  AWTP (2007) scenario is: "the heat
generated in the lower chromosphere is absorbed by strong
radiative loss, some excess heating occurs in the upper
chromosphere and lower transition region, which drives
chromospheric evaporation and gives rise to filled coronal loops".

This is an attempt by AWTP (2007) to provide a mechanism for  the
hot \ {\it chromospheric evaporation}\ . In this process they seem
to forget about the \ {\it upflows} \ or, perhaps,  for them the \
{\it evaporation} \ constitutes the {\it upflow}\ ! Without paying
due attention to the interaction of the flows and the magnetic
field, they seem to miss that when upflows (whatever their initial
temperature) enter the closed magnetic field region, heating will
always take place due to the dissipation of short--scale flow
energy. The heating due to flow energy dissipation may, in
addition, be augmented by  secondary heating. The term "secondary
heating" in this case has a very different connotation -- it bears
no relation to the AWTP (2007) \ {\it excess heating} \ prior to \
{\it filling} \ (loop formation). In our model, the "secondary
heating" may occur to simply sustain (against, say, radiation
losses) the hot bright loop. The emerging scenario, then, is not
{\it the filling} of some hypothetical virtual loop with hot gas.
The loop, in fact, is created by the interaction of the flow and
the ambient filed; its formation and heating are simultaneous and
the "loop" has no ontological priority to the flow. [see
\cite{MMNS2} simulation results where the radiative losses are
taken into account].

\vspace{1cm}

\noindent{\bf Acknowledgements}

Authors are grateful for the hospitality at the Abdus Salam
International Centre for Theoretical Physics, where this work was
done. The second author acknowledges the support from Abdus Salam
ICTP Regular Associateship award and Georgian National Science
Foundation grant 69/07 (GNSF/ST06/4-057).


\clearpage

\begin{thebibliography}{99}

\bibitem[Aschwanden {\it et al.} 2007]{AWTP} Aschwanden, M.J.,
Winebarger, A., Tsiklauri, D. and Peter, H. 2007, \apj, 659, 1673.

\bibitem[Aschwanden {\it et al.} 2001a]{A1} Aschwanden, M.J., Poland
A.I., and Rabin D.M. 2001a, Ann. Rev. Astron. Astrophys., 39, 175.

\bibitem[Aschwanden 2001b]{A2} Aschwanden, M.J. 2001b, \apj, 560,
1035.

\bibitem[Kjeldseth-Moe \& Brekke 1998]{KB} Kjeldseth-Moe, O. and Brekke,
B. 1998, Solar Phys., 182, 73.

\bibitem[Klimchuk 2006]{klimchuk} Klimchuk, J. A. 2006, Solar
Phys., 234, 41.

\bibitem[Lopez Fuentes \& Klimchuk \& Mandrini 2007]{klimchuk2} Lopez
Fuentes, M.C. Klimchuk, J. A. and Mandrini, C.H. 2007, \apj, 657,
1127.

\bibitem[Mahajan {\it et al.} 1999]{MMNS1} Mahajan, S.M., Miklaszewski,
R., Nikol'skaya, K.I., and Shatashvili N.L. February 1999,
Preprint IFSR \#857, {\it The University of Texas, Austin}, 67
pages.

\bibitem[Mahajan {\it et al.} 2001]{MMNS2}  Mahajan, S.M., Miklaszewski,
R., Nikol'skaya, K.I., and Shatashvili, N.L. 2001, Phys. Plasmas, 8,
1340.

\bibitem[Mahajan {\it et al.} 2002a]{mnsy} Mahajan, S.M.,
Nikol'skaya, K.I., Shatashvili, N.L. \& Yoshida, Y. 2002a, \apj,
576, L161.

\bibitem[Mahajan {\it et al.} 2005]{RD} Mahajan, .M., Shatashvili, N.L.,
Mikeladze, S.V. and Sigua, K.I. 2005, \apj, 634, 419.

\bibitem[Mahajan {\it et al.} 2006]{msms} Mahajan, .M., Shatashvili, N.L.,
Mikeladze, S.V. and Sigua, K.I. 2006, Phys. Plasmas, 13, 062902.

\bibitem[Mahajan \& Yoshida 1998]{MY-1} Mahajan, S.M., and Yoshida,
Z. 1998, Phys. Rev. Lett., 81, 4863.

\bibitem[Mahajan {\it et al.} 2002b]{channel1} Mahajan, S.M., Miklaszewski, R.,
Nikol'skaya, K.I. and Shatashvili,N.L.. 2002b, Adv. Space Res.
30(3), 545.

\bibitem[Mahajan {\it et al.} 2003]{channel2} Mahajan, S.M., Miklaszewski, R.,
Nikol'skaya, K.I. and Shatashvili,N.L..  ArXiv: astro-ph/0308012
(2003)

\bibitem[Ohsaki {\it et al.} 2001]{osym1} Ohsaki, S., Shatashvili, N.L.,
Yoshida, Z., and Mahajan, S.M. 2001, \apj, 559, L61.

\bibitem[Ohsaki {\it et al.} 2002]{osym2} Ohsaki, S., Shatashvili, N.L.,
Yoshida, Z., and Mahajan, S.M. 2002, \apj, 570.

\bibitem[Schrijver {\it et al.} 1999]{schrijver}  Schrijver, C.J.,
Title, A.M., Berger, T.E., Fletcher, L., Hurlburt, N.E., Nightingale,
R.W., Shine, R.A., Tarbell, T.D., Wolfson, J., Golub, L., Bookbinder,
J.A., Deluca, E.E., McMullen, R.A., Warren, H.P., Kankelborg, C.C.,
Handy B.N., and DePontieu, B. 1999, Solar Phys., 187, 261.

\bibitem[Seaton {\it et al.} 2001]{ami1} Seaton, D.B., Winebarger, A.R.,
DeLuca, E.E., Golub, L., and Reeves, K.K. 2001, \apj, 563, L173.

\bibitem[Warren \& Winebarger 2007]{warren} Warren, H. P. and Winebarger, A.R. 2007, \apj,
666, 1245.

\bibitem[Wilhelm 2001]{wilhelm} Wilhelm, K. 2001, Astrophys. and
Astronomy, 360, 351.

\bibitem[Winebarger \& DeLuca \& Golub 2001]{ami} Winebarger, A.M.,
DeLuca, E.E., and Golub, L. 2001, \apj, 553, L81.

\bibitem[Winebarger {\it et al.} 2002]{ami3} Winebarger, A.R., Warren,
H., Van Ballagooijen, A., DeLuca E.E., and Golub, L. 2002, \apj, 567,
L89.

\end{thebibliography}
\end{document}